\documentstyle[prd,aps,epsf,eqsecnum]{revtex}
\begin{document}
\def\d{{\mathrm{d}}}
\def\nbox#1#2{\vcenter{\hrule \hbox{\vrule height#2in
\kern#1in \vrule} \hrule}}
\def\sq{\,\raise.5pt\hbox{$\nbox{.09}{.09}$}\,}
\def\sqb{\,\raise.5pt\hbox{$\overline{\nbox{.09}{.09}}$}\,}    
\newcommand{\tab}{$\langle T_{ab} \rangle$}

\title
{Linear Response and the Validity of the Semi-Classical
Approximation in Gravity}

\author
{
Paul R. Anderson$^{1}$
\thanks{electronic address: anderson@wfu.edu},
Carmen Molina-Par\'{\i}s$^{2,3}$
\thanks{electronic address: molina@maths.warwick.ac.uk},
and
Emil Mottola$^{2}$
\thanks{electronic address: emil@lanl.gov}
}

\address{
$^{1}$
Department of Physics,
Wake Forest University, Winston-Salem,
North Carolina, 27109\\
$^{2}$
T-8, Theoretical Division, Los Alamos
National Laboratory, Los Alamos, New Mexico, 87545
\\
$^{3}$
Mathematics Institute, University of Warwick,
Coventry CV4 7AL, UK}

\date{LA-UR-02-2072}
\maketitle

\abstract {We propose a quantitative test for the validity of the
semi-classical approximation in gravity, namely that the solutions
to the semi-classical equations should be stable to linearized
perturbations, in the sense that no gauge invariant perturbation should
become unbounded in time. We show that a self-consistent linear response
analysis of these perturbations based upon an invariant
effective action principle involves metric fluctuations about
the
mean semi-classical geometry and brings in the two-point
correlation function of the quantum energy-momentum tensor in a
natural way. The properties of this correlation function are
discussed and it is shown on general grounds that it contains no
state-dependent divergences and requires no new renormalization  
counterterms beyond those required in the leading order
semi-classical approximation.}

\section{Introduction}

There are many well known difficulties that arise when attempting to
combine quantum field theory and general relativity into a full
quantum theory of gravity. Almost certainly, a consistent quantum
theory at the Planck scale requires a fundamentally different set of
principles from those of classical general relativity, in which even
the concept of spacetime itself is likely to be radically altered.
Yet over a very wide range of distance scales, from that of the
electroweak interactions ($10^{-16}$ cm) to cosmology ($10^{27}$ cm),
the basic framework of a spacetime metric theory obeying general
coordinate invariance is assumed to be valid, and receives
phenomenological support both from the successes of flat space
quantum field theory at the lower end of this scale and classical
general relativity at its upper end. Hence whatever the full quantum
theory of gravity entails, it should reduce to an effective low
energy field theory on this very broad range of some $33$ orders of
magnitude of distance~\cite{hu,donoghue}.

To the extent that quantum effects are relevant at all in
gravitational phenomena within this range of scales one would expect
to be able to apply {\it semi}-classical techniques to the low energy
effective theory.  For the purposes of this paper the quantization of
matter fields in curved spacetime backgrounds is what we mean by the
semi-classical approximation to gravity. Thus the spacetime metric
$g_{ab}$ is treated as a classical $c$-number field and its quantum
fluctuations are neglected. An interesting question then arises: Can a
quantitative criterion for the validity of this semi-classical
approach be given?  One would like to have some test of its validity,
preferably within the semi-classical framework itself.

To be clear we should distinguish what we mean in this paper by the
semi-classical approximation to gravity from the ordinary loop
expansion, which is sometimes also called semi-classical. In the ordinary loop
expansion of the effective action $\hbar$ is the formal loop expansion
parameter.  As a result both the matter and gravitational fields are treated
on exactly the same footing. However the technical issues involved in even
defining a one-loop effective action for gravitons that respects both
linearized gauge and background field coordinate invariance are difficult
enough to have impeded progress in the standard loop expansion in
gravity~\cite{dewitt,dw-mp}. An unambiguous definition of the corresponding
conserved and gauge invariant energy-momentum tensor for gravitons on an
arbitrary curved spacetime has not yet been given.  Apart from such technical
shortcomings an ordinary loop expansion is physically ill-suited to many
applications that have been and are likely to be of interest in semi-classical
gravity, such as particle creation in the early universe, or black hole
radiance, where quantum effects significantly affect the background
geometry after some period of time. This is because when quantum
effects significantly affect the classical geometry the loop
expansion breaks down. 

The semi-classical approximation to gravity we discuss in this paper treats
the matter fields as quantum but the spacetime metric as classical. This
asymmetric treatment can be justified formally by replicating the number of
matter fields $N$ times and taking the large $N$ limit of the quantum
effective action for the matter fields in an arbitrary background metric
$g_{ab}$~\cite{tomboulis}.  Since no assumption of the weakness or
perturbative nature of the metric is assumed, the large $N$ expansion is able
to address problems in which gravitational effects on the matter are strong
and the matter fields can have a significant effect on the classical geometry
in turn. The absence of quantum gravitational effects in the lowerst order
large $N$ approximation also means that most of the technical obstacles
arising from the quantum fluctuations of the geometry are avoided. General
coordinate invariance is assured, provided only that the matter effective
action is regularized and renormalized in a manner which respects this
invariance~\cite{dewitt}. In that case the quantum expectation value of the
matter energy-momentum tensor $\langle T_{ab}\rangle$ is necessarily
conserved.

Assuming that the classical energy-momentum
tensor for the matter field(s) vanishes (an assumption that may easily be
relaxed if necessary), the unrenormalized form of the semi-classical
backreaction equations take the form
\begin{equation}
G_{ab} + \Lambda g_{ab} = 8\pi G_{_N} \langle T_{ab}\rangle \; .
\label{eq:scE}
\end{equation}
Here $G_{ab}$ is the Einstein tensor, $\Lambda$ is the cosmological
constant (which may be taken to be zero in some applications),
$G_{_N}$ is Newton's constant, and {\tab} is the expectation value of
the energy-momentum tensor operator of the quantized matter field(s).
Among the technical issues that must be confronted is the
renormalization of the expectation value of $T_{ab}$, a quartically
divergent composite operator in $d=4$ spacetime dimensions. Hence
renormalization of its expectation value requires the introduction of
fourth order counterterms in the effective (low energy) action,
modifying the geometric terms on the left side of
eq. (\ref{eq:scE})~\cite{birrell-davies}.  This renormalization of
very high momenta, ultraviolet divergences must be performed in a way
that is consistent with general coordinate invariance and leaves
undisturbed the vanishing of the covariant divergence of both sides of
eq. (\ref{eq:scE}).

Once a renormalized semi-classical theory has been defined, one possible route
to investigating its validity is to compare calculations in a theory of
quantum gravity with similar semi-classical calculations. Since a full quantum
theory is lacking, this has been done only in some simplified models of
quantum gravity. Large quantum gravity effects were found in
three-dimensional models by Ashtekar~\cite{ashtekar} and 
Beetle~\cite{beetle}. In four dimensions Ford considered the case of
graviton production in a linearized theory
of quantum gravity on a
flat space background and compared the results with the
production of
gravitational waves in semi-classical gravity~\cite{ford}.  He
found
that they were comparable when the renormalized energy-momentum
(connected) correlation function,
\begin{equation}
\langle T_{ab}(x) T_{cd}(y)  \rangle_{\rm con} \equiv
\langle T_{ab}(x) T_{cd}(y) \rangle - \langle  T_{ab}(x) \rangle 
\langle T_{cd}(y) \rangle   
\label{eq:TT}
\end{equation}
satisfied the condition
\begin{equation} 
\langle T_{ab}(x) T_{cd}(y) \rangle_{\rm con} \ll 
\langle
T_{ab}(x)  \rangle \langle T_{cd}(y)\rangle 
\; . 
\label{eq:ford} 
\end{equation}

The limits of validity of the semi-classical approximation have also
been studied without making reference to a specific model of
quantum gravity. Kuo and Ford~\cite{kuo-ford} proposed that a measure of how strongly
the semi-classical approximation is violated can be given by how
large the quantity,
\begin{equation} 
\Delta_{abcd} (x,y) \equiv
\left|\frac{\langle T_{ab}(x) T_{cd}(y) \rangle_{\rm con}} {\langle
T_{ab}(x) T_{cd}(y)\rangle}\right|
\label{eq:k-f}
\end{equation}
is, where it is assumed that the expectation values in this
expression are suitably renormalized. They suggested that the
expectation values be normal ordered, which is strictly correct only
in a flat spacetime. Later authors used other renormalization
schemes. It is important to note that eq. (\ref{eq:k-f}) is
coordinate dependent, since both the numerator and denominator are
{\it tensor} quantities. The situation is complicated further by the
regularization and renormalization issues that arise in defining the
quantities appearing in this expression. Kuo and Ford~\cite{kuo-ford} 
computed
the quantity 
\begin{equation} 
\Delta (x) \equiv \left|\frac{\langle T_{00}(x)
T_{00}(x) \rangle_{\rm con}} {\langle T_{00}(x)
T_{00}(x)\rangle}\right| \label{eq:variance}
\end{equation}
for a free scalar field in flat space for several states including the Casimir
vacuum.  They found that it vanishes in a coherent state, whereas in many
other cases, including the Casimir vacuum, it is of order unity.

Wu and Ford~\cite{wu-ford-1} computed the radial flux component of eq. 
(\ref{eq:k-f}) in the case of an evaporating black hole far from the event
horizon.  They found that it was of order unity over time scales comparable to
the black hole mass, but that it averages to zero over much larger
times. In
a normal ordering prescription they found state
dependent divergent terms, which should not be
present if all
divergences are local and therefore absorbable into purely
geometric
UV counterterms.
 
Phillips and Hu~\cite{phillips-hu-1} used zeta function
regularization to compute $\Delta (x)$ with the denominator replaced
by the quantity $\langle T_{00}(x)\rangle^2$, for a free scalar
field in some curved spacetimes having Euclidean sections. They also
computed $\Delta (x)$ for a scalar field in
flat space in the
Minkowski vacuum state using both point splitting
 and a smearing
operator to remove the divergences~\cite{phillips-hu-2}.  
 For
the
flat space calculation they found that $\Delta (x)$ depends on
the
 direction the points are split, but that it is of order unity
regardless of how the points are split.  They used their results to
criticize the Kuo-Ford conjecture and to suggest that the criteria
for the validity of the semi-classical approximation should depend
on
 the scale at which the system is being probed.
 
Although it is somewhat unclear what the dimensionless small parameter which
controls the inequality (\ref{eq:ford}) is, Ford's initial work and these
subsequent discussions draw attention to the importance of the higher point
correlation functions of the energy-momentum tensor.  It is quite clear in
qualitative terms that if the higher point {\it connected} correlation
functions of $T_{ab}$ are large (in an appropriate sense to be determined), it
cannot be correct to neglect them completely, as the semi-classical equations
(\ref{eq:scE}) certainly do.

The energy-momentum correlation function $\langle T_{ab}(x) T_{cd}(y)
\rangle_{\rm con}$ has been computed in some cases.  Carlitz and
Willey~\cite{carlitz-willey} computed it for a scalar field in a two
dimensional spacetime with a moving boundary.  Roura and
Verdaguer~\cite{roura-verdaguer} give its expression for a massless minimally
coupled scalar field in de Sitter spacetime in the case that the points are
spacelike separated and geodesically connected.  Wu and Ford~\cite{wu-ford-2}
showed that in the simple case of radiation exerting a force on a mirror, the
quantum fluctuations in the radiation pressure are due to a
state-dependent cross term in the energy-momentum correlation
function.

In the same paper~\cite{wu-ford-2} Wu and Ford also addressed the Kuo-Ford
conjecture and the above mentioned criticism of it by Phillips and Hu.  They
stated that the conjecture is incomplete because it does not address the
effect of divergent state-dependent terms. They suggested that any
criterion for the validity of the semi-classical approximation should
be a non-local one that involves integrals over the world lines of
test particles. They also argued that the question of whether the
semi-classical approximation is valid depends on the specifics of a
given situation including the scales being probed and the choice of
initial quantum state. Although technical problems  such as
renormalization and coordinate invariance complicate matters, this
body of previous work suggests that the correlation function $\langle
T_{ab}(x) T_{cd}(y) \rangle_{\rm con}$ should play an important role
in determining the validity of the  semi-classical approximation.
However, the proper context for incorporating and making use of the
information contained in this correlation function remains somewhat
in question.

In this paper we propose an unambiguous quantitative criterion for the
validity of the semi-classical approximation, namely that solutions
to the semi-classical equations should be stable against linearized
perturbations. According to standard linear response
theory~\cite{mar-sch} we show that the linearized equations for the
perturbed metric depend on the two-point correlation function of the
energy-momentum tensor evaluated in the semi-classical metric
$g_{ab}$.  The correlation function in this case consists of the
retarded commutator of two energy-momentum tensor operators, and
hence the perturbations are manifestly causal. Moreover the UV
divergences in this correlation function are exactly those required
to renormalize the semi-classical theory itself. This ensures that no
state dependent divergences occur.  Finally, gauge dependence of the
solutions to the linear response equations is easily studied within
the framework of linearized coordinate transformations of the
semi-classical background, so that ambiguities related to quantities
such as (\ref{eq:k-f}) do not arise.

Another advantage of this criterion for the validity of the
semiclassical approximation is that it lies strictly within the
context of that approximation. In this way one avoids problems
such as gauge invariance of the energy-momentum tensor for gravitons,
that inevitably appear if one tries to go beyond the semi-classical
approximation and include quantum effects due to the
gravitational field.

To understand qualitatively the role of the two-point correlation
function in the validity of the semi-classical approximation, it is
helpful to consider the physical analogy between semi-classical
gravity and semi-classical electromagnetism.  The connected
correlation function (\ref{eq:TT}) measures the gravitational vacuum
polarization which contributes to the proper self-energy of the
linearized graviton fluctuations around the background metric, just
as the current two-point correlation function, $\langle j^a(x)
j^b(y)\rangle$, measures the electromagnetic vacuum polarization
which contributes to the proper self-energy of the
photon~\cite{CHKM}. Hence if these polarization effects are
significant, the gravitational fluctuations of the metric must be
taken into account in some form and the semi-classical approximation,
with or without backreaction, has certainly broken down, at least in
the form specified by eq. (\ref{eq:scE}), where all fluctuations of
the metric have been ignored.  Moreover, in quantum electromagnetism
we know exactly how to take these fluctuation effects into account,
namely by scattering and interaction Feynman diagrams involving the
photon propagator~\cite{CHKM}. These processes are important not only
in scattering between a few particles at high energies but also in
low energy processes in hot or dense plasmas.  Analogous statements
should be applicable to gravity whenever quantum correlations are
important.  Thus, if the linear response validity criterion is not
satisfied, then there will be no avoiding the technical difficulties
of quantizing the gravitational field, {\it even if we seek to  
understand only the infrared behavior of a semi-classical
approximation to the effective theory far below the Planck scale.}

In the next section the properties
of the large $N$ semi-classical approximation in gravity and its
renormalization within the covariant effective action framework are
reviewed. In Section III the linear response theory for the
semi-classical backreaction equations is introduced.  The form of the
two-point correlation function for the energy-momentum tensor that
appears in the linear response equations is given and its properties
and renormalization are discussed. Then our proposal for a necessary
condition for the validity of the semi-classical approximation is
presented. Finally some possible applications of our
criterion to the study of quantum effects in cosmological and black
hole spacetimes are suggested.

\section{Semi-Classical Gravity and Renormalization}

The most direct route to the semi-classical equations (\ref{eq:scE}) is via
the effective action method in the large $N$ limit. We consider the specific
example of $N$ non-interacting scalar fields.  Generalizations to interacting
fields and fields of other spin are straightforward, but as they are not
required to expose the main elements of the stability criterion, we
shall treat only this simplest case in detail.

The classical action for one scalar field is
\begin{equation} 
S_{\rm m}[\Phi, g]
= -\frac{1}{2}\int \d^4 x \;  \sqrt{-g}\, 
[ (\nabla_a \Phi) g^{ab} (\nabla_b \Phi) +m^2 \Phi^2+ \xi R\Phi^2] 
\; ,
\end{equation}
where $\nabla_a$ denotes the covariant derivative in the metric $g_{ab}$,
$\xi$ the dimensionless curvature coupling, and $R$ the scalar curvature.  The
path integral over the free scalar field $\Phi$ is Gaussian and may be
performed formally by inspection, {\it i.e.,}
\begin{equation}
\int[{\cal D}\Phi]\exp \left({i\over \hbar}S_{\rm m}[\Phi, g]\right)
= \exp\left(-{1\over 2} {\rm Tr}
\log G^{-1}[g]\right) \equiv 
\exp \left({i\over \hbar}S_{\rm eff}^{(1)}[g]\right)
\; ,
\label{eq:intM}
\end{equation}
where
\begin{equation}
G^{-1}[g] \equiv -\sq + m^2 + \xi R 
\end{equation}
is the inverse propagator of the scalar field in the background metric
$g_{ab}$, and the (generally non-local) functional
\begin{equation}
S_{\rm eff}^{(1)} [g]= {i\hbar \over 2}\,{\rm Tr} \log G^{-1}[g]
\label{eq:effM}
\end{equation}
may be regarded as the effective action in this
metric. It contains an explicit factor of $\hbar$ and records the
quantum effects of the free scalar field in the
arbitrary curved background metric $g_{ab}$.  No assumption about
smallness of the metric deviations from flat spacetime or any other
preferred spacetime has been made.

The expectation value of the energy-momentum tensor of the quantum matter
field in this background can be formally obtained by the variation
\begin{equation}
\langle T_{ab}\rangle 
= -{2\over \sqrt{-g}} {\delta\over \delta g^{ab}} S_{\rm eff}^{(1)}[g]
\; .
\label{eq:Texp}
\end{equation}
By Noether's theorem, this (unrenormalized) expectation value is covariantly
conserved, provided that the effective action $S_{\rm eff}^{(1)}[g]$ is
invariant under general coordinate transformations. However, $\langle
T_{ab}\rangle $ is divergent and requires a careful UV regularization and
renormalization procedure consistent with this
invariance~\cite{birrell-davies}. 

In physical terms the UV regularization and renormalization mean that the
theory may not be trustworthy at infinitely short time and distance scales,
but that the lack of information about the physics at those small scales may
be absorbed into a finite number of parameters in the effective low energy
theory at much larger scales. Since the effective lagrangian and
energy-momentum tensor have canonical scale dimension $d$ (in $d$ spacetime
dimensions), the number of parameters is given by the number of local
coordinate invariant scalars up to dimension $d$ (in $d$ dimensions).  In
$d=4$ dimensions these are the parameters of the Einstein-Hilbert action plus
the coefficients of the two independent fourth order invariants $R^2$ and
$C_{abcd}C^{abcd}$, where $R$ is the scalar curvature and $C_{abcd}$ is the
Weyl tensor.  Thus we shall consider the total low energy effective
gravitational action
\begin{equation}
S_{\rm eff}[g] = S_{\rm eff}^{(1)}[g] + {1\over 16\pi G_{N}}\int \, 
\d^4x\, \sqrt{-g}\, (R-2\Lambda) 
+ {1\over 2} \int\, \d^4x\, \sqrt{-g} \, 
\left(\alpha R^2 + \beta C_{abcd} C^{abcd} \right)
\; ,
\label{eq:effS}
\end{equation}
with arbitrary dimensionless constants $\alpha$ and $\beta$.  Renormalization
means that $G_N$, $\Lambda$, $\alpha$, and $\beta$ may be chosen to depend
on the UV cutoff (introduced to regulate the divergences in the one-loop term
$S_{\rm eff}^{(1)}[g]$) in such a way as to cancel those divergences and
render the total $S_{\rm eff}[g]$ independent of the cutoff. Hence the four
parameters of the local geometric terms (up to fourth order derivatives of the
metric which are {\it a priori} independent of $\hbar$) must be considered as
parameters of the same order as the corresponding divergent terms in 
$S_{\rm  eff}^{(1)}[g]$, which from (\ref{eq:effM}) is first order in $\hbar$.
Formally this may be justified by considering $N$ identical copies of the
matter field, so that $S_{\rm eff}^{(1)}[g]$ is replaced by 
$NS_{\rm eff}^{(1)}[g]$ and $G_{N}^{-1}$, $\Lambda/G_{N}$,
$\alpha$, and $\beta$ are rescaled by a factor of $N$. In this way
all the terms in eq. (\ref{eq:effS}) are now of the same order in
$N$ as $N \rightarrow \infty$.

This formal rescaling by $N$ is carried out at the level of the
generating functional $S_{\rm eff}[g]$ of connected $n$-point vertices (which
are the {\it inverse} of $n$-point Green's functions) rather than the Green's
functions themselves. Therefore, it has the net effect of resumming
the quantum effects contained in the one-loop diagrams of the matter
field(s) to all orders in the metric $g_{ab}$.  The large $N$
expansion and its relationship to the standard loop expansion have
been extensively studied in both $\Phi^4$ theory and electrodynamics
(both scalar and spinor QED) in flat space~\cite{CHKM}.  The QED case
is most analogous to the present discussion with the classical vector
potential $A_{\mu}$ replaced by the metric $g_{ab}$. The large $N$
approximation (\ref{eq:effS}) is also invariant under changes in the
ultraviolet renormalization scale (by definition of the UV cutoff
dependence of the local counterterms), and is equivalent to the UV
renormalization group (RG) improved one-loop approximation.

In the present case of a free field theory in curved spacetime, it is the
large $N$, RG improved one-loop approximation that is necessary to derive the
semi-classical equations (\ref{eq:scE}) with backreaction, for only in such a
resummed loop expansion can the one-loop quantum effects of $\langle
T_{ab}\rangle$ influence the nominally classical background metric $g_{ab}$.
As mentioned in the previous section, in the ordinary (unimproved)
loop expansion the quantum fluctuations of the matter can make at
most small corrections to the background metric. The large $N$
approximation also explicitly preserves the covariance properties of
the theory, since it can be derived from an invariant action
functional (\ref{eq:effS}). The divergences in $\langle T_{ab}
\rangle $ are in one-to-one correspondence with the local
counterterms in the action $S_{\rm   eff} [g]$, whose variations with
respect to $g_{ab}$ produce, in addition to the terms in the
classical Einstein equations, the fourth order tensors
\begin{mathletters} 
\begin{eqnarray} 
^{(1)}H_{ab}&&\equiv {1\over\sqrt{-g}}{\delta\over\delta g^{ab}} 
\int\, \d^4\,x\,\sqrt{-g}\, R^2 =
2 g_{ab}\sq R  -2\nabla_a\nabla_b R + 2 R R_{ab} - {g_{ab}\over 2}R^2 
\; ,  
\\
^{(C)}H_{ab}&&\equiv {1\over \sqrt{-g}}{\delta\over\delta g^{ab}}
\int\, \d^4\,x\,\sqrt{-g}\, C_{abcd}C^{abcd} = 4\nabla^c \nabla^d C_{acbd}
+ 2R^{cd}C_{acbd}
\; .
\end{eqnarray}
\label{eq:Hdef}
\end{mathletters}

\noindent
Hence the
variation of the effective action (\ref{eq:effS}) gives the equations
of motion for the spacetime metric for zero expectation value of the
free scalar field $\Phi$
\begin{equation} 
\alpha\ ^{(1)}H_{ab} +
\beta\ ^{(C)}H_{ab} + {1\over 8\pi G_{_N}}\left(G_{ab} + \Lambda
g_{ab}\right) =  \langle T_{ab}\rangle_{_R} \; ,  
\label{eq:scF}  
\end{equation}
where $\langle T_{ab}\rangle_{_R}$ is the renormalized expectation
value of the energy-momentum tensor of the scalar field.
Non-zero values of $\langle \Phi\rangle$ and self-interactions are
easily taken into account in the large $N$ approximation
as well~\cite{CHKM}. 

It is worth emphasizing that the UV renormalization of the energy-momentum
tensor and the covariant form of the equations of motion (\ref{eq:scF}) are
justified by formal appeal to an underlying covariant action principle
(\ref{eq:effS}) whose variation they are.  Although particular regularization
and renormalization procedures, such as non-covariant point-splitting or
adiabatic subtraction, may break explicit covariance, the result must be of
the form (\ref{eq:scF}) with a covariantly conserved $\langle
T_{ab}\rangle_{_R}$ or the procedure does not correspond to the
addition of local counterterms up to dimension $d=4$ in the effective
action, as required by the general principles of renormalization
theory. Thus the renormalization of the effective action
(\ref{eq:effS}) suffices in principle to renormalize the equations of
motion and its higher variations, a fact we make use of in the next
section.

In all cases the large $N$ approximation is equivalent to a Gaussian path
integration for the quantum matter fields, in which the spacetime metric and
gravitational degrees of freedom have been treated as $c$-numbers, coupled
only to the expectation value of the energy-momentum tensor through
(\ref{eq:scE}). This energy-momentum tensor expectation value can be expressed
as a coincident limit of local derivatives of the one-loop matter Green's
function $G[g](x,x)$ in the background metric $g_{ab}$.  That is, it requires
solving the differential equation $G^{-1}[g]\circ G[g] = 1$ or more explicitly
\begin{equation}
\left(-\sq + m^2 + \xi R\right) G[g](x, x') 
= {\delta^4(x,x')\over \sqrt{-g}} \; , 
\label{eq:grfn}
\end{equation}
concurrently with the semi-classical backreaction equation
(\ref{eq:scF}). It is the exact solution of this equation without any
perturbative re-expansion of $G[g]$, and the
resulting self-consistent solution of eq. (\ref{eq:scF}) for the
metric $g_{ab}$ that constitutes the principal non-perturbative RG
improved feature of the large $N$ limit.

The equations of motion (\ref{eq:scF}) which are
the original eqs. (\ref{eq:scE}) modified by the additional terms
required by the UV renormalization of $\langle T_{ab}\rangle$ are
fourth order in derivatives of the metric.
As is known from the general theory of differential equations, if the
order of the equations is changed by adding higher derivative terms,
the solutions of the modified equations fall into two classes, {\it
viz.,} those that approach the solutions of the lower order equations
as $\alpha, \beta \rightarrow 0$, and those which are singular in
that limit. The latter class of solutions are not present in the
lower order theory and physically correspond to those solutions which
vary on Planck length and time scales (in order for the higher
derivative terms to be of the same order as the lower derivative
terms). Such solutions are clearly not trustworthy in a low energy 
effective theory which breaks down at the Planck
scale.
Furthermore, we are not interested in the stability issue 
 raised by
the presence of such rapid variations in
 time and space. Quite to
the contrary, we would like to study the validity of the
semi-classical approximation at scales far below the Planck scale,
where this approximation is at least not invalid {\it a priori}.
 In
this regime the effects of the higher order local
 terms in eq.
(\ref{eq:scF}) can be handled by an approach
 similar to that
proposed by Simon~\cite{simon} and Parker and
 Simon~\cite{p-s}.
 
\section{Linear Response and the Stability Criterion}

Just as the first variation of the action functional
(including that for the matter field) with respect to $g_{ab}$ gives
Euler-Lagrange equations for the mean value of the metric, its second
variation gives information about the fluctuations of the metric
about its mean value. The second variation of the action functional is
equivalent to the first variation of the semi-classical
Einstein equations, namely   
\begin{eqnarray} &&
\delta \left[
\alpha\ ^{(1)}H_{ab} +
\beta\ ^{(C)}H_{ab} + {1\over 8\pi G_{_N}}\left(G_{ab} + \Lambda
g_{ab}\right) 
\right] = 
\delta \langle T_{ab}\rangle_{_R}
\nonumber\\ 
&& \qquad 
= -\int \d^4x'\sqrt{-g'}\,
\Pi_{ab}^{({\rm ret})
\, c'd'}(x, x')\,\delta g_{c'd'}(x')
\; ,
\label{eq:linres}
\end{eqnarray}
where
\begin{equation}
\Pi_{ab}^{({\rm ret})\, c'd'}(x, x') \equiv -i \theta (t, t') 
\left\langle \left[T_{ab}(x), 
T^{c'd'}(x')\right]_-\right\rangle 
= - {\delta^2  S_{\rm eff}^{(1)}[g]\over \delta g^{ab}(x)
\delta g_{c'd'}(x')}
\label{eq:var}
\end{equation}
is the retarded gravitational polarization operator of the matter fields in
the curved background $g_{ab}$. This retarded operator corresponds to the
causal boundary conditions of the effective action functional 
$S_{\rm eff}^{(1)}[g]$, which are explicitly enforced by the Schwinger-Keldysh
closed time path (CTP) contour of the time integration~\cite{CTP}. The
linearized fluctuation $\delta g_{ab}(x)$ obeys an integro-differential
equation (\ref{eq:linres}) in which the integral depends only on the past of
$x$, due to the causal boundary conditions, and which involves the
two-point correlation function of the matter energy-momentum tensor. 
According to the general principles of linear response analysis, this
retarded correlation function is evaluated in the background geometry
of the leading order solution of the semi-classical equations
(\ref{eq:scF}). 

Before discussing the stability criterion we make some
additional technical remarks. First, the polarization operator
$\Pi_{ab}^{({\rm ret})\, c'd'}(x, x')$ has local divergences involving
derivatives of $\delta^{(4)} (x, x')$ when the spacetime points $x$
and $x'$ coincide. Each of these divergences is proportional to one
of the local tensor variations on the left side of eq. (\ref{eq:linres}). 
To see this note that if the UV renormalization is  performed at the
level of the effective action (\ref{eq:effS}) then all variations of
the renormalized effective action are necessarily finite. Hence all
the local UV divergences may be removed by adjusting the coefficients
$\alpha$, $\beta$, $G_{_N}$, and $\Lambda$ in the effective action,
and there are no state dependent divergences in either $\langle
T_{ab}\rangle_{_R}$ or $\Pi_{ab}^{({\rm ret})\, c'd'}(x, x')$.

Next, since the polarization operator is determined by the second
variation of the same effective action that determines the
energy-momentum tensor, it also obeys the covariant conservation law
\begin{equation} 
\nabla^a \Pi_{ab}^{({\rm ret})\, c'd'}(x, x') 
= \nabla_{c'} \Pi_{ab}^{({\rm ret})\, c'd'}(x, x') = 0
\; .
\end{equation}
 
Finally, the equations (\ref{eq:linres}) are covariant in form and
therefore are non-unique up to linearized coordinate (gauge)
transformations,
\begin{equation}
\delta g_{ab} \rightarrow \delta g_{ab} + \nabla_a X_b + \nabla_b X_a
\; ,
\label{eq:gauge}
\end{equation}
for any vector field $X_a$. Singular gauge transformations in the initial 
data for $\delta g_{ab}$ are certainly not allowed, and some care is
required to decide whether time dependent linearized gauge
transformations which grow in time without bound are allowed or not.
Since the action principle is fundamental to the present approach,
any transformation of the form (\ref{eq:gauge}) for which the action
(\ref{eq:effS}) is not invariant (due to boundary or surface terms)
is not a true invariance and should be excluded from the set of
allowable gauge transformations of the linear response equations
(\ref{eq:linres}).

We are now in a position to state our stability criterion for the
the semi-classical approximation. A necessary condition for the
validity of the large
 $N$ semi-classical equations of motion
(\ref{eq:scF}) is that the
 linear response equations
(\ref{eq:linres}) should have no solutions
 with finite non-singular
initial data for which any linearized gauge
 invariant scalar
quantity grows without bound.
 Such a quantity must be constructed
only from the linearized
 metric perturbation $\delta g_{ab}$ and its
derivatives, and it
 must be invariant under allowed gauge
transformations of
 the kind described by eq. (\ref{eq:gauge}). 
 
The existence of any solutions to the linear response equations with
unbounded growth in time, that cannot be removed by an allowed
linearized gauge transformation (\ref{eq:gauge}), implies that
the influence of the growing gravitational fluctuations on the
semi-classical background geometry are large and must be taken into
account in the evolution of the background itself. That is to say, if
the gravitational fluctuations around the background grow, even
if they were initially small, then the leading order
semi-classical equations (\ref{eq:scF}), which neglect these
fluctuations, must eventually break down.

To this point in time a consistent inclusion of gravitational
fluctuations in the dynamical evolution of the background metric has
not been attempted, even in the most symmetric and cosmologically
relevant of cases, namely the Robertson-Walker background. However,
if the linearized solutions show any growing modes then such
self-consistent inclusion of the effect of the gravitational
fluctuations beyond the leading order semi-classical approximation
would be required.

If the linear response equation (\ref{eq:linres}) has physical
unstable solutions at space and time scales determined by the
semi-classical background metric, which are very far from the Planck
scale, then the analysis should be reliable. An instability in the low
energy infrared effective theory means that it is the semi-classical
solution to eq. (\ref{eq:scF}) that must be modified by taking these
{\it infrared} gravitational fluctuations into account
self-consistently, rather than abandoning the entire framework of a
spacetime metric description of gravity.

One important application of this criterion is to de Sitter spacetime. 
A number of different arguments leads to the conclusion that de
Sitter spacetime is not the ground state of a quantum theory of
gravity with a cosmological term~\cite{mot-des}. In fact, the
two-point correlation function of the energy-momentum tensor for a
scalar field was estimated in~\cite{mot-fd} and argued to contribute
to a gauge invariant growing mode on the horizon time scale.  This
proposition could be tested by detailed calculation of the two-point
correlation function of the energy-momentum tensor and the solutions
of the linear response equations (\ref{eq:linres}).

A second relevant application of the criterion is to black hole
spacetimes. Ever since the discovery of black hole radiance it has
been recognized that the quantum behavior of black holes is
qualitatively different from the classical analogs at {\it long}
times, since semi-classical black holes decay at late
times while classical black holes are absolutely stable. In the
Hartle-Hawking state~\cite{harhaw} one can construct a static solution
to the semi-classical equations (\ref{eq:scF}) that is quite close to
the classical one near the horizon~\cite{york,hky,ahwy}.  However, the
stability of this self-consistent solution has not been investigated.
The validity criterion proposed in this paper provides a clear test
for the stability of the self-consistent solutions in both the
black hole and de Sitter cases which we plan to investigate in future
publications.

\acknowledgements  
P.\ R.\ A.\ would like to thank T-8, Los Alamos National Laboratory
for its hospitality, as well as J. Donoghue,
L. Ford, B.L. Hu, and N. Phillips for helpful conversations.  This
work was supported in part by grant numbers PHY-9800971 and
PHY-0070981 from the National Science Foundation. It was also
supported in part by contract number W-7405-ENG-36 from the Department
of Energy.


\begin{references}

\bibitem{hu} 
B.L. Hu, Physica A {\bf 158}, 399 (1989).

\bibitem{donoghue} 
J.F. Donoghue, Phys. Rev. D {\bf 50}, 3874 (1994).

\bibitem{dewitt}
B.S. DeWitt,
{\it  in Les Houches 1985, Proceedings, Architecture Fundamental Interactions 
at Short Distances, Vol. 2, 1023-1057.} 

\bibitem{dw-mp}
B.S. DeWitt and C. Molina-Par\'\i s,
Mod. Phys. Lett. A {\bf 13}, 2475 (1998).

\bibitem{tomboulis}
E. Tomboulis, Phys. Lett. B {\bf 70}, 361 (1977).

\bibitem{birrell-davies}
N.D. Birrell and P.C.W. Davies,
{\it Quantum Fields in Curved Space}
(Cambridge University Press, Cambridge, England, 1982),
and references therein.

 
\bibitem{ashtekar}  
A. Ashtekar,  
Phys. Rev. Lett. {\bf 77}, 4864 (1996). 
 
\bibitem{beetle}  
C. Beetle, Adv. Theor. Math. Phys. {\bf 2}, 471 (1998). 

\bibitem{ford}
L.H. Ford, Ann. Phys. (N.Y.) {\bf 144}, 238 (1982). 

\bibitem{kuo-ford} 
C.-I. Kuo and L.H. Ford, 
Phys. Rev. D {\bf 47}, 4510 (1993).

\bibitem{wu-ford-1}
C.-H. Wu and L.H. Ford, 
Phys. Rev. D {\bf 60}, 104013 (1999).

\bibitem{phillips-hu-1} 
N.G. Phillips and B.L. Hu, 
Phys. Rev. D {\bf 55}, 6123 (1997).

\bibitem{phillips-hu-2} 
B.L. Hu and N.G. Phillips, 
Int. J. Theor. Phys. {\bf 39}, 1817 (2000);\hfill\break
N.G. Phillips and B.L. Hu, Phys. Rev. D {\bf 62}, 084017 (2000).

\bibitem{carlitz-willey} 
R.D. Carlitz and R.S. Willey, 
Phys. Rev. D {\bf 36}, 2327 (1987).

\bibitem{roura-verdaguer} 
A. Roura and E. Verdaguer, 
Int. J. Theor. Phys. {\bf 38}, 3123 (1999).

\bibitem{wu-ford-2} 
C.-H. Wu and L.H. Ford, Phys. Rev. D {\bf 64}, 045010 (2001);
e- print gr-qc/0102063.

\bibitem{mar-sch}
P.C. Martin and J. Schwinger, Phys. Rev. {\bf 115}, 1342 (1959).

\bibitem{CHKM}
F. Cooper, S. Habib, Y. Kluger, E. Mottola, J. P. Paz,
and P.R. Anderson, Phys. Rev. D {\bf 50}, 2848 (1994);\hfill\break
F. Cooper, S. Habib, Y. Kluger, and E. Mottola, Phys. Rev.
D {\bf 55}, 6471 (1997).

\bibitem{simon} 
J.Z. Simon, Phys. Rev. D {\bf 41}, 3720 (1990); 
{\it ibid.} 
{\bf 43}, 3308 (1991).

\bibitem{p-s} 
L. Parker and J.Z. Simon, Phys. Rev. D {\bf 47}, 1339 (1993).

\bibitem{CTP}
J. Schwinger, J. Math. Phys. {\bf 2}, 407 (1961);\hfill\break
L.V. Keldysh, Zh. Eksp. Teor. Fiz. {\bf 47}, 1515 (1964)
[Sov. Phys. JETP {\bf 20}, 1018 (1965)];\hfill\break
K.-C. Chou, Z.-B. Su, B.-L. Hao, and L. Yu,
Phys. Rep. {\bf 118}, 1 (1985).

\bibitem{mot-des}
E. Mottola, Phys. Rev. D {\bf 31}, 754 (1985);
{\it ibid.} {\bf 33}, 1616 (1986);\hfill\break
P.O. Mazur and E. Mottola, Nucl. Phys. B {\bf 278}, 694 (1986);\hfill\break
I. Antoniadis and E. Mottola, J. Math. Phys. {\bf 32}, 1037 (1991);
\hfill\break
E. Mottola, J. Math. Phys. {\bf 36}, 2470 (1995).

\bibitem{mot-fd}
E. Mottola, Phys. Rev. D {\bf 33}, 2136 (1986).

\bibitem{harhaw}
J. B. Hartle and S. W. Hawking, Phys. Rev. D {\bf 13}, 2188 (1976)

\bibitem{york} 
J.W. York, Jr., Phys. Rev. D {\bf 31}, 775 (1985).

\bibitem{hky} 
D. Hochberg, T.W. Kephart, and J.W. York, Jr., 
Phys. Rev. D {\bf 48}, 479 (1993).

\bibitem{ahwy} 
P.R. Anderson, W.A. Hiscock, J. Whitesell, and J.W. York Jr.,
Phys. Rev. D {\bf 50}, 6427 (1994).

\end{references}
\end{document}